\documentclass[aps,prb,twocolumn,superscriptaddress]{revtex4}
\usepackage{mathrsfs}
\usepackage{epsfig}
\usepackage{graphicx}
\usepackage{amsfonts}
\usepackage[figuresright]{rotating}
\usepackage{amssymb}
\usepackage{amsmath}
\usepackage{dcolumn}
\usepackage{bm}
\usepackage{color}

\def\be{\begin{equation}} \def\ee{\end{equation}}
\def\bea{\begin{eqnarray}} \def\eea{\end{eqnarray}}

\newcommand{\WQC} {Wilczek Quantum Center, School of Physics and Astronomy, Shanghai Jiao Tong University, Shanghai 200240, China}

\newcommand{\SQC} {Key Laboratory of Artificial Structures and Quantum Control, School of Physics and Astronomy, Shanghai Jiao Tong University, Shanghai 200240, China}

\newcommand{\Pittsburgh} {Department of Physics and Astronomy, University of Pittsburgh, Pittsburgh, Pennsylvania 15260, USA}

\newcommand{\SRCQC}{Shanghai Research Center for Quantum Sciences, Shanghai 201315, China}

\newcommand{\zhiyuan} {Zhiyuan College, Shanghai Jiao Tong University, Shanghai 200240, China}

\begin{document}
\title{Order by disorder in frustration-free systems: a quantum Monte Carlo study of a two-dimensional PXP model}

\author{Mingxi Yue}
\thanks{These authors contributed equally to this work.}
\affiliation{\WQC}

\author{Zijian Wang}
\thanks{These authors contributed equally to this work.}
\affiliation{\WQC}
\affiliation{\zhiyuan}

\author{Bhaskar Mukherjee}
\affiliation{\WQC}
\affiliation{\Pittsburgh}

\author{Zi Cai}
\email{zcai@sjtu.edu.cn}
\affiliation{\WQC}
\affiliation{\SQC}
\affiliation{\SRCQC}

\begin{abstract}  
In this study, an order by disorder mechanism has been proposed in a two-dimensional PXP model, where the extensive degeneracy of the classical ground-state manifold is due to  strict occupation constraints  instead of geometrical frustrations. By performing an unbias large-scale quantum Monte Carlo simulation, we find that local quantum fluctuations, which usually work against long-range ordering, lift the macroscopic classical degeneracy and give rise to a compressible ground state with charge-density-wave long-range order. A simple trial wavefunction has been proposed to capture the essence of the ground-state of the two-dimensional  PXP model.  The finite temperature properties of this model have also been studied, and we find a thermal phase transition with an universality class of two-dimensional Ising model.

\end{abstract}


\maketitle

{\it Introduction --} Order by disorder mechanism is one of the most remarkable phenomena in geometrically frustrated materials,  where it is impossible to  simultaneously  minimize interacting energy for all bonds in the system. Hence extensive degeneracy arises in the classical ground-state manifold.  Such a classical ground-state degeneracy may be partially or completely lifted by subtle effects that normally work against ordering in a process termed ``order by disorder'', where particular configurations are selected out of the degenerate manifold  and long-range order is  restored\cite{Villain1979,Henley1987,Henley1989}. For instance,  thermal fluctuations can give rises to entropic differences between configurations, hence degeneracy-breaking free-energy terms at finite temperature\cite{Villain1979,Henley1987}, while at zero temperature, taking into account  quantum fluctuations may leads to degenerate-breaking zero point energy that favors ordered states\cite{Henley1989,Chubukov1992}. Up to now, most studies in this field  concentrate on the frustrated systems, a question thus arises: whether frustration is a necessity for us to observe the order by disorder phenomenon?

In this paper,  we attempt to answer this question, focusing for simplicity on a two-dimensional(2D) PXP model on a square lattice. In spite of its extreme simplicity, this model does not only provide a prototypical example of order by disorder phenomena in frustration-free system, but is also closely related with recent progress with quantum simulators based on Rydberg atomic systems\cite{Endres2016,Barredo2016,Kumar2018},  where a neutral atom can strongly interact with its neighbors via a Rydberg  blockade mechanism  which prevents two nearby atoms from being simultaneously excited into the excited states\cite{Jaksch2000}.  In such a highly tunable and controllable platform,  the interplay between constraints resulting from the Rydberg blockade and the quantum fluctuations  could give rise to exotic quantum matters\cite{Laycock2011,Lesanovsky2012,Labuhn2016,Leseleuc2019,Samajdar2020} as well as non-equilibrium quantum many-body dynamics\cite{Bernien2017,Keesling2019,Turner2019,Ho2019}.

By performing a numerically exact quantum Monte Carlo (QMC) simulation\cite{Prokofev1998}, we study the ground-state of a PXP model in a square lattice, which can be realized in an interacting Rydberg atomic system with zero detuning\cite{Lesanovsky2012b}. It is shown that in the absence of laser driving, any classical configuration satisfying the  Rydberg blockade condition is a ground-state of this model, yielding a macroscopic degeneracy in the ground-state manifold. For each atom in the lattice, the driving laser field induces coupling between the ground state and Rydberg state, thus acts as a local quantum fluctuation. Counter-intuitively,  such an off-diagonal term (in the Fock basis),  which usually works against  long-range ordering, is found to lift the macroscopic classical degeneracy, and give rise to a charge-density-wave state with a spontaneous $Z_2$ symmetry breaking. This result has provided a different perspective view of the order by disorder phenomenon, where the extensive degeneracy lifted by quantum fluctuations comes from occupation constraints instead of geometrical frustration.

{\it Model and method --} In terms of the hard-core boson language, the Hamiltonian of a PXP model reads
\begin{equation}
H=\Omega \sum_{\mathbf{i}}(\tilde{b}_\mathbf{i}+\tilde{b}_\mathbf{i}^\dag) \label{eq:PXP}
\end{equation}
where $\tilde{b}_\mathbf{i}$ ($\tilde{b}_\mathbf{i}^\dag$) is the annihilation (creation) operator of the hard-core bosons operating on the constraint Hilbert space $\mathbb{H}$ that excludes those configurations with two bosons located on neighboring sites. Similar models have also been proposed to describe correlated bosonic systems in tiled optical lattice\cite{Sachdev2002,Sengupta2004,Pielawa2011}.  In terms of  spin-$\frac 12$ operators, Eq.(\ref{eq:PXP}) describes a PXP model with the Hamiltonian $H_{PXP}=\Omega\sum_\mathbf{i}  \tilde{\sigma}_\mathbf{i}^x$. Compare to the one dimensional version of the related models whose ground state and non-equilibrium properties have been intensively studied\cite{Olmos2012,Olmos2012b,Turner2019,Ho2019,Lin2019,Mark2020}, the 2D PXP model is much less studied\cite{Michailidis2020,Michailidis2020b}.

The PXP model can also be realized in strongly interacting Rydberg atomic system subjected to a coherent laser driving field, whose Hamiltonian reads as follows:
\begin{equation}
H_{R}=\sum_\mathbf{i} [\Omega (|g\rangle_\mathbf{i} \langle e|+|e\rangle_\mathbf{i} \langle g|)-\delta |e\rangle_\mathbf{i} \langle e|]+V\sum_{\langle \mathbf{ij}\rangle} \hat{n}^e_\mathbf{i} \hat{n}^e_\mathbf{j} \label{eq:Ryd}
\end{equation}
where $\langle\mathbf{ij}\rangle$ indicates a pair of adjacent lattice sites in the $L\times L$ square lattice. $|g\rangle_i$ ($|e\rangle_i$) denote the internal atomic ground (excited) state of the atom on site $\mathbf{i}$, and $\hat{n}_\mathbf{i}^e=|e\rangle_\mathbf{i}\langle e|$ is the density operator of the excited state on site $\mathbf{i}$.  $\Omega$ and $\delta$ are the Rabi frequency and the detuning of the coherent laser driving field, and $V$ is the strength of the interaction between the Rydberg atoms.  Here we only consider the nearest neighbor interaction, which can be experimentally realized by properly tuning the Rydberg blockade radius. The ground-state phase diagram of Eq.(\ref{eq:Ryd}) has been studied by density matrix renormalization group (DMRG) method, where different crystalline ground states and corresponding quantum phase transitions have been found\cite{Samajdar2020}. Here, to illustrate its relationship with the order by disorder phenomena, we consider Eq.(\ref{eq:Ryd}) in the limit $V\rightarrow \infty$ and $\delta\rightarrow 0$, where Eq.(\ref{eq:Ryd}) turns to the 2D PXP model.(\ref{eq:PXP}).

Throughout this paper, we study the 2D PXP model described by Eq.(\ref{eq:PXP}) using a continuous-time QMC algorithm, which allows us to study both the ground state and thermal properties up to a significantly large system size. Notice that local quantum fluctuation in  Eq.(\ref{eq:PXP}) does not result in  sign problem in the Monte Carlo sampling, nor does the occupation constraint, thus the QMC simulation is unbias and  numerically exact. To sample those (imaginary)time-space configurations according to their weights in the partition function, we choose an ergodic set of updates consists of randomly inserting/removing onsite  pair vertices in the configurations\cite{Greitemann2018}. The constraint complicates the simulation by introducing correlations between adjacent sites. To approach the ground state,  in the QMC simulations, we scale the inverse temperature as $\beta=L$  and take the thermodynamic limit $L\rightarrow \infty$.

 \begin{figure}[htb]
\includegraphics[width=0.99\linewidth,bb=94 60 750 552]{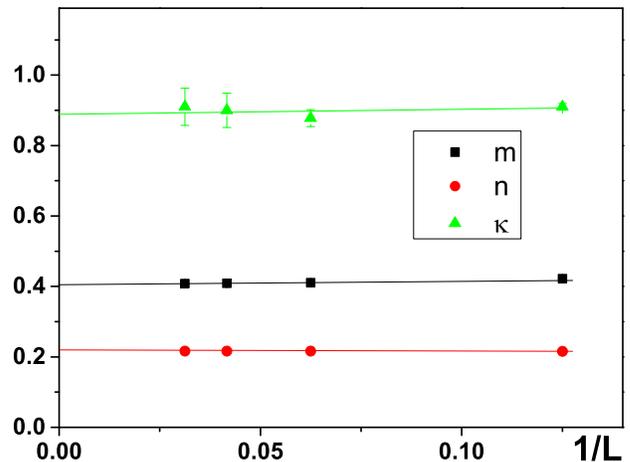}
\caption{(Color online) Finite size scaling of  the CDW order parameter $m$, density of the bosons $n$ and the compressibility $\kappa$ in the 2D $L\times L$ PXP model obtained by QMC simulations with $\beta=L$. } \label{fig:fig1}
\end{figure}
{\it Classical limit --} We first focus on the case without quantum fluctuations ($\Omega=0$), where any classical configuration in the Fock basis satisfying the occupation constraint has the same energy $E=0$. In an one-dimensional(1D) lattice, the number of these states  diverges exponentially with the system size L as  $\sim e^{L\ln\alpha}$ with $\alpha=(\sqrt{5}+1)/2$\cite{Domb1960}. In a 2D square lattice, the dimension of this constraint Hilbert space $\mathbb{H}$ still diverges exponentially with the number of lattice sites $N=L^2$, while it is difficult to find an analytical expression as simple as that in the 1D case.

{\it Order by disorder in the ground state --} Now we consider the effect of the local quantum fluctuation in Eq.(\ref{eq:PXP}), which breaks the particle number conservation.  We first focus on the ground state properties,  the finite size scaling of the CDW order parameter
\begin{equation}
m=\frac 2N\sqrt{\langle[\sum_\mathbf{i} (-1)^{i_x+i_y} \hat n_\mathbf{i}]^2\rangle}\label{eq:CDWm}
\end{equation}
as well as the density of the bosons
\begin{equation}
n=\frac 1N\sum_\mathbf{i} \langle n_\mathbf{i}\rangle
\end{equation}
are plotted in Fig.\ref{fig:fig1}, from which we can find a non-vanishing CDW order parameter $m_0=0.403(6)$ at the density $n_0=0.221(5)$ in the thermodynamic limit. These results indicates that the local quantum fluctuation, which usually works against long-range ordering, select a quantum state with CDW long-range order from the extensively degenerate classical ground-state manifold. Notice that the physical quantities of this model barely depend on the system size $L$, which indicates that a feature of locality. In another word, the ground state might be close to a product state.  More information of the ground state can be obtained from the compressibility, which is defined based on the total number fluctuation as:
\begin{equation}
\kappa=\frac{\beta}{L^2}(\langle \hat{N}^2\rangle-\langle \hat{N}\rangle^2),
\end{equation}
where $\hat{N}=\sum_\mathbf{i}\hat{n}_\mathbf{i}$ is the total particle number operator of the hard-core bosons. The finite size scaling of $\kappa$ is also plotted in Fig.\ref{fig:fig1}, which indicates that the ground state is compressible. However, due to the explicit breaking of the U(1) symmetry in Eq.(\ref{eq:PXP}), its ground state is not a superfluid.

The order by disorder phenomenon in this frustration-free system and emergent CDW order can be understood as a consequence of competition between the quantum fluctuation and occupation constraint. The quantum fluctuations (spin flip terms $\tilde{\sigma}^x_i$ in the spin-$\frac 12$ language) could further lower the energy of the system by resonating the degenerate classical configurations, thus are  energetically favored in the ground state. However, two ``spin-flip'' operators acting on a pair of adjacent sites will inevitably give rise to the configurations that violate the occupation constraint. As a compromise, the system favors a ground state with ``spin-flips'' as many as possible, while they try to keep away from each other  to avoid Rydberg blockade. One of the simplest example satisfying these requirements is a product state with a wavefunction: $\Phi_A=\prod_i |\phi\rangle_i$, where $|\phi\rangle_i=|0\rangle_i$ if $i\in A$, and $|\phi\rangle_i=\frac 1{\sqrt{2}}(|0\rangle_i-|1\rangle_i)$ if $i\in B$ with A(B) indicating the A(B) sublattice in the square lattice. $|0\rangle_i$ ($|1\rangle_i$) indicates the ground (excited) state of the Rydberg atoms on site i.   The state $\Phi_A$ as well as its sublattice symmetric counterpart $\Phi_B=\Phi_{A\rightarrow B}$ spontaneously break the translational symmetry, thus exhibits a long rang order which can qualitatively explain the observed CDW order in the ground state of Eq.(\ref{eq:PXP}).

Since there is no spontaneous symmetry breaking for any finite system, we propose a trial wavefunction
\begin{equation}
\Phi_{trial}=\frac{1}{\sqrt{2}}(\Phi_A+\Phi_B)\label{eq:trial}
\end{equation}
 which is an equally-weighted superposition of the symmetry breaking states to restore the symmetry but keep the long-range CDW correlations.  In the Fock basis, the  wavefunction $\Phi_A$ ( $\Phi_B$)  indicate  an equally-weighted superposition of all the possible classical configurations in  sublattice A(B), while leave the other sublattice empty. A similar trial wavefunction without translational symmetry breaking has been proposed for the 1D PXP model\cite{Ovchinnikov2003}.

 \begin{figure}[htb]
\includegraphics[width=0.99\linewidth]{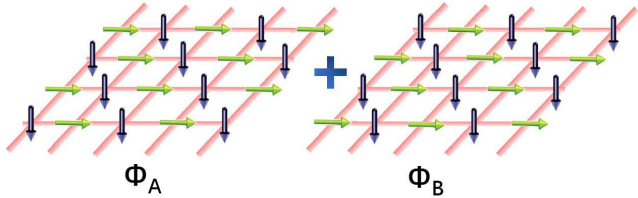}
\caption{(Color online) Schematic diagram of the trial wavefunction in Eq.(\ref{eq:trial}) in terms of spin language. }\label{fig:pic}
\end{figure}

\begin{table}[htb]
\begin{tabular}{|p{2.5cm}<{\centering}|p{1.9cm}<{\centering}|p{1.9cm}<{\centering}|p{1.9cm}<{\centering}|}
\hline
 & $n$ & m & $e_g$ \\ \hline
trial  & 0.25&  $0.5$  &  $-0.5\Omega$\\ \hline
Lanczos($6\times6$)  & 0.209 & 0.417 & $-0.513\Omega$  \\ \hline
QMC ($L\rightarrow \infty$) & 0.221 & 0.403  &  $-0.512\Omega$   \\ \hline
\end{tabular}
\caption{Comparisons of physical quantities (particle density n, CDW order parameter m, and average energy $e_g=E_g/N$) of the ground state  obtained by the trial wavefunction, Lanczos method for a $6\times 6$ system and the QMC simulations.} \label{Table:1}
\end{table}

Such a trial wavefunction in Eq.(\ref{eq:trial}), in spite of its extreme simplicity, can qualitatively capture most features of the ground state of Eq.(\ref{eq:PXP}). To verify this point  numerically, we  compare the results predicted by this trial wavefunction to our numerical results. We first focus on the physical quantities, for instance the average energy predicted by the trial wavefunction .(\ref{eq:trial}) is $-0.5\Omega$, which is only $2.4\%$ higher than its exact value. In addition, the particle density predicted by wavefunction (\ref{eq:trial}) is $0.25$, which is also not very far from its exact value $0.22$. To directly compare the trial wavefunction to the exact one, we perform the exact diagonalization to obtain the exact ground-state wavefuction $|\Phi_E\rangle$. Notice that the occupation constraint significantly reduces the dimension of the Hamiltonian matrix, thus enables us to obtain $|\Phi_E\rangle$  by Lanczos algorithm for a small system.  Take a $6\times 6$ system for an example, the dimension of the constraint Hilbert space $\mathbb{H}$ is  only $0.0035\%$ of that of the full Hilbert space, and the overlap between the the exact wavefunction and the trial wavefunction.(\ref{eq:trial}) $\langle \Phi_{trial}|\Phi_E\rangle=0.84888$, which is pretty high considering the facts that the dimension of these two wavefunctions are 2406862, and there are no any tunable parameter in the trial wavefunction.(\ref{eq:trial})

 \begin{figure}[htb]
\includegraphics[width=0.9\linewidth,bb=123 106 977 800]{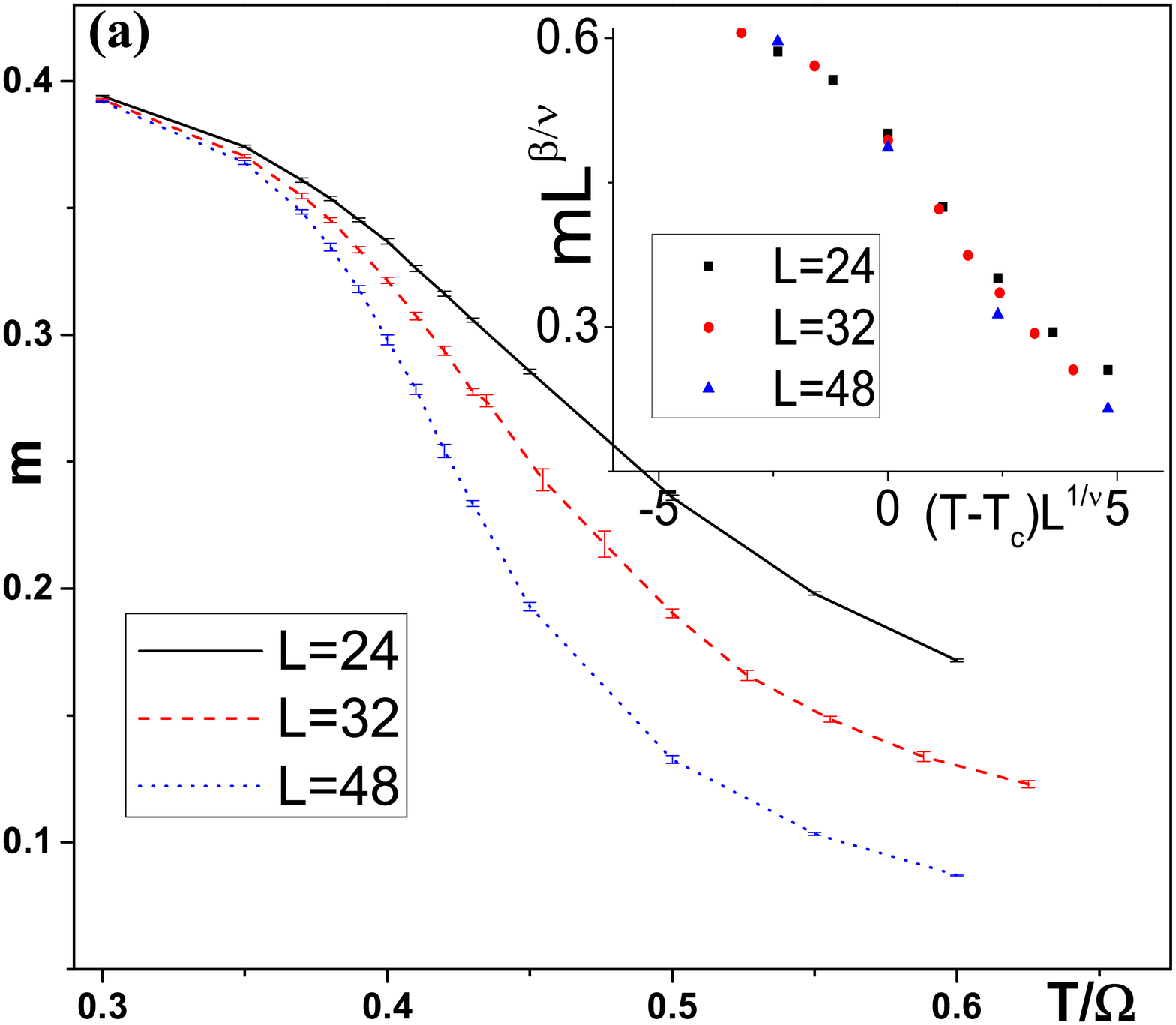}
\includegraphics[width=0.9\linewidth,bb=123 106 977 800]{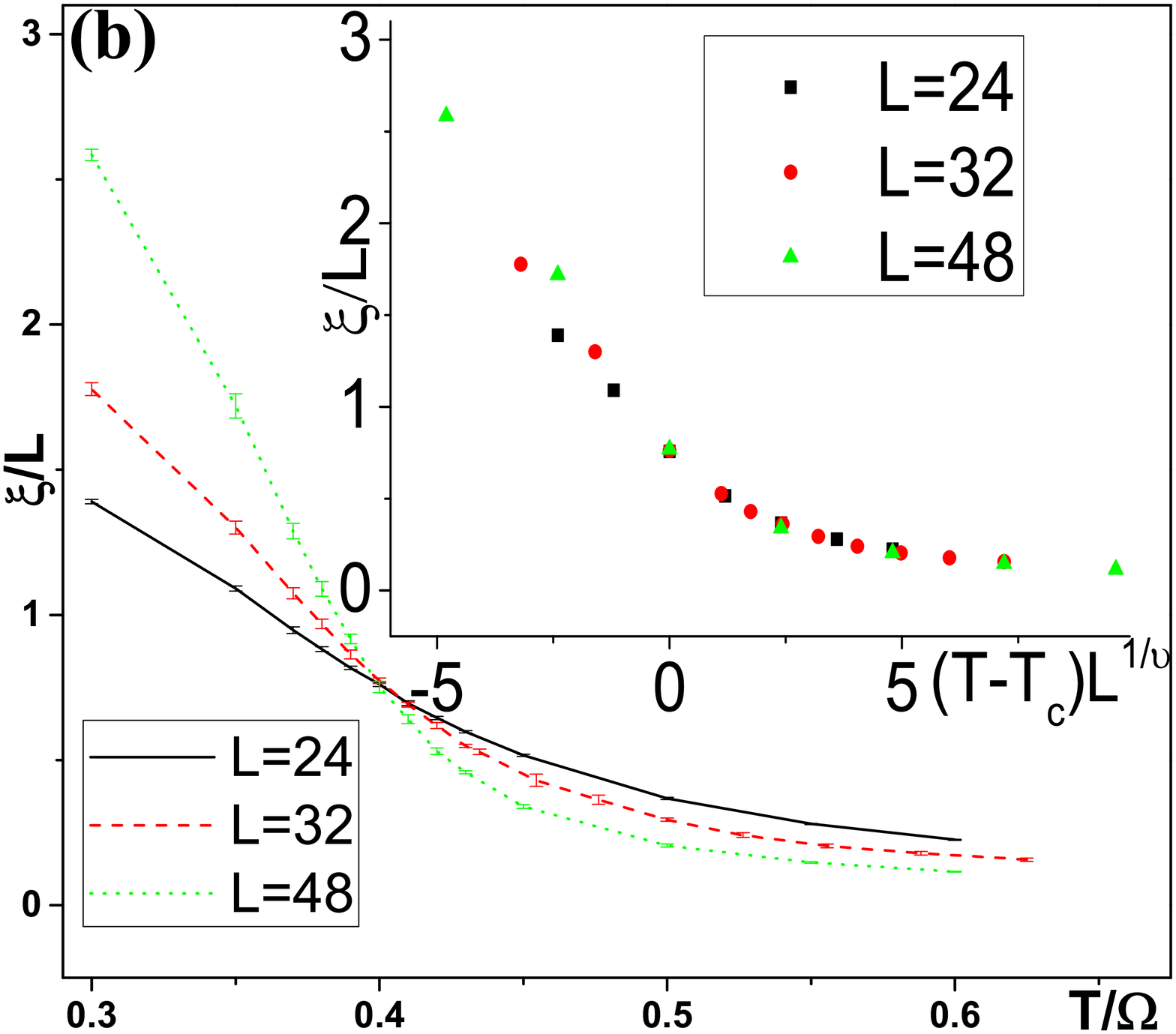}
\caption{(Color online) (a) CDW order parameter m and (b) renormalized correlation length $\xi/L$ as a function of temperature T for different system size $L$ in the QMC simulation. The insets are the data collapse for (a) the order parameter and (b) the renormalized correlation length using the critical exponents $\nu=1$ and $\beta=1/8$.
} \label{fig:fig2}
\end{figure}
{\it Finite temperature phase transition  --}  It has been shown that the ground state of the 2D PXP Eq.(\ref{eq:PXP}) spontaneously  breaks the  $Z_2$ translational symmetry and exhibit the CDW long range order. Considering the fact that there is only one parameters $\Omega$ in PXP Eq.(\ref{eq:PXP}), at zero temperature, this CDW order will persist irrespective of the strength of the quantum fluctuation $\Omega$. At finite temperature,  one may expect a competition between quantum and thermal fluctuations, which may result in a finite-temperature phase transition. To verify this point numerically, we calculate CDW order parameter as a function of temperature, as shown in Fig.\ref{fig:fig2}(a).  We also calculate the CDW correlation length $\xi$ (along x-direction) in the QMC simulation as\cite{Sandvik2010}:
\begin{equation}
\xi=\frac 1{q_1}\sqrt{\frac{S(\pi,\pi)}{S(\pi+q_1,\pi)}-1}
\end{equation}
where $S(\mathbf{Q})=\frac{1}{N^2} \sum_{ij} e^{i\mathbf{Q}\cdot(\mathbf{i}-\mathbf{j})}\langle n_\mathbf{i} n_\mathbf{j}\rangle$ is the structure factor of the density-density correlation. $q_1=\frac{2\pi}L$. The CDW order parameter $m$ is defined the same as Eq.(\ref{eq:CDWm}). The normalized correlation length $\xi/L$  as a function of temperature for systems with different sizes $L$ has been plotted in Fig.\ref{fig:fig2} (b), from which we can find a scaling invariant point indicating a critical point at $T_c=0.4\Omega$. Near the critical point, the correlation length diverges as $\xi\sim |T-T_c|^{-\nu}$, and the order parameter scales as $m\sim |T-T_c|^\beta$.  To explore the universal class of this phase transition, we study the scaling behavior around the critical point. The data collapses as shown in the insets of Fig.\ref{fig:fig2} indicate that  the critical exponents $\beta=1/8$ and $\nu=1$, which agrees with those of the 2D Ising universal class.

{\it Conclusion and outlook --} In summary, we observe a order by disorder phenomenon in a frustration-free system, where the extensive classical ground state degeneracy is due to occupation constraint instead of geometrical frustration. It is shown that in a 2D PXP model, the competition between local quantum fluctuation and occupation constraint gives rise to a compressible ground state with  CDW long-range order, which could be destroyed by thermal fluctuation via a finite-temperature phase transition with an universality class of 2D Ising model.

Future developments will include a generalization of our method and result to systems with different types of CDW orders and quantum fluctuations.  For instance, Samajdar {\it et al} have proposed complex CDW orders in 2D interacting Rydberg atomic systems\cite{Samajdar2020}. Even though most of them are due to the direct interactions between the Rydberg atoms in the classical limit, there could be some regimes in the global phase diagram whose CDW orders are induced by the order-by-disorder mechanism. Different types of quantum fluctuations,  especially those preserving the total particle number($\tilde{b}_i^\dag \tilde{b}_j$ for an instance), are of particular interest, and  one may expect a CDW-to-superfluid transition with increasing doping in this case. In addition, the groundstate phase diagram of the PXP models with additional terms (e.g. the chemical potential $-\mu \tilde{n}_i$) and the corresponding quantum critical behavior can also been studied\cite{Samajdar2020}, even though this term with $\mu\neq 0$ lifts the extensively degeneracy of the corresponding classical model ($\Omega=0$) thus is not directly related with the order-by-disorder phenomena discussed here.    Furthermore, our QMC algorithm is ready to be generalized to frustrated systems (e.g. kagome and triangle lattices), where the interplay between the frustration, quantum fluctuations and constraint may give rise to exotic quantum many-body states\cite{Samajdare2021,Verresen2021}.

{\it Acknowledgments}.---This work is supported by the National Key Research and Development Program of China (Grant No. 2020YFA0309000 and No. 2016YFA0302001), NSFC of  China (Grant No.11574200),  Shanghai Municipal Science and Technology Major Project (Grant No.2019SHZDZX01).


\begin{thebibliography}{34}
\expandafter\ifx\csname natexlab\endcsname\relax\def\natexlab#1{#1}\fi
\expandafter\ifx\csname bibnamefont\endcsname\relax
  \def\bibnamefont#1{#1}\fi
\expandafter\ifx\csname bibfnamefont\endcsname\relax
  \def\bibfnamefont#1{#1}\fi
\expandafter\ifx\csname citenamefont\endcsname\relax
  \def\citenamefont#1{#1}\fi
\expandafter\ifx\csname url\endcsname\relax
  \def\url#1{\texttt{#1}}\fi
\expandafter\ifx\csname urlprefix\endcsname\relax\def\urlprefix{URL }\fi
\providecommand{\bibinfo}[2]{#2}
\providecommand{\eprint}[2][]{\url{#2}}

\bibitem[{\citenamefont{Villain}(1979)}]{Villain1979}
\bibinfo{author}{\bibfnamefont{J.}~\bibnamefont{Villain}}, \bibinfo{journal}{Z.
  Phys. B} \textbf{\bibinfo{volume}{33}}, \bibinfo{pages}{31}
  (\bibinfo{year}{1979}).

\bibitem[{\citenamefont{Henley}(1997)}]{Henley1987}
\bibinfo{author}{\bibfnamefont{C.~L.} \bibnamefont{Henley}},
  \bibinfo{journal}{J. Appl. Phys.} \textbf{\bibinfo{volume}{61}},
  \bibinfo{pages}{3962} (\bibinfo{year}{1997}).

\bibitem[{\citenamefont{Henley}(1989)}]{Henley1989}
\bibinfo{author}{\bibfnamefont{C.~L.} \bibnamefont{Henley}},
  \bibinfo{journal}{Phys. Rev. Lett.} \textbf{\bibinfo{volume}{62}},
  \bibinfo{pages}{2056} (\bibinfo{year}{1989}).

\bibitem[{\citenamefont{Chubukov}(1992)}]{Chubukov1992}
\bibinfo{author}{\bibfnamefont{A.}~\bibnamefont{Chubukov}},
  \bibinfo{journal}{Phys. Rev. Lett.} \textbf{\bibinfo{volume}{69}},
  \bibinfo{pages}{832} (\bibinfo{year}{1992}).

\bibitem[{\citenamefont{Endres et~al.}(2016)\citenamefont{Endres, Bernien,
  Keesling, Levine, Anschuetz, Krajenbrink, Senko, Vuletic, Greiner, and
  Lukin}}]{Endres2016}
\bibinfo{author}{\bibfnamefont{M.}~\bibnamefont{Endres}},
  \bibinfo{author}{\bibfnamefont{H.}~\bibnamefont{Bernien}},
  \bibinfo{author}{\bibfnamefont{A.}~\bibnamefont{Keesling}},
  \bibinfo{author}{\bibfnamefont{H.}~\bibnamefont{Levine}},
  \bibinfo{author}{\bibfnamefont{E.~R.} \bibnamefont{Anschuetz}},
  \bibinfo{author}{\bibfnamefont{A.}~\bibnamefont{Krajenbrink}},
  \bibinfo{author}{\bibfnamefont{C.}~\bibnamefont{Senko}},
  \bibinfo{author}{\bibfnamefont{V.}~\bibnamefont{Vuletic}},
  \bibinfo{author}{\bibfnamefont{M.}~\bibnamefont{Greiner}}, \bibnamefont{and}
  \bibinfo{author}{\bibfnamefont{M.~D.} \bibnamefont{Lukin}},
  \bibinfo{journal}{Science} \textbf{\bibinfo{volume}{354}},
  \bibinfo{pages}{1024} (\bibinfo{year}{2016}).

\bibitem[{\citenamefont{Barredo et~al.}(2016)\citenamefont{Barredo, Leseleuc,
  Lienhard, Lahaye, and Browaeys}}]{Barredo2016}
\bibinfo{author}{\bibfnamefont{D.}~\bibnamefont{Barredo}},
  \bibinfo{author}{\bibfnamefont{S.~D.} \bibnamefont{Leseleuc}},
  \bibinfo{author}{\bibfnamefont{V.}~\bibnamefont{Lienhard}},
  \bibinfo{author}{\bibfnamefont{T.}~\bibnamefont{Lahaye}}, \bibnamefont{and}
  \bibinfo{author}{\bibfnamefont{A.}~\bibnamefont{Browaeys}},
  \bibinfo{journal}{Science} \textbf{\bibinfo{volume}{354}},
  \bibinfo{pages}{1021} (\bibinfo{year}{2016}).

\bibitem[{\citenamefont{Kumar et~al.}(2018)\citenamefont{Kumar, Wu, Giraldo,
  and Weiss}}]{Kumar2018}
\bibinfo{author}{\bibfnamefont{A.}~\bibnamefont{Kumar}},
  \bibinfo{author}{\bibfnamefont{T.-Y.} \bibnamefont{Wu}},
  \bibinfo{author}{\bibfnamefont{F.}~\bibnamefont{Giraldo}}, \bibnamefont{and}
  \bibinfo{author}{\bibfnamefont{D.~S.} \bibnamefont{Weiss}},
  \bibinfo{journal}{Nature} \textbf{\bibinfo{volume}{561}}, \bibinfo{pages}{83}
  (\bibinfo{year}{2018}).

\bibitem[{\citenamefont{Jaksch et~al.}(2000)\citenamefont{Jaksch, Cirac,
  Zoller, Rolston, C\^ot\'e, and Lukin}}]{Jaksch2000}
\bibinfo{author}{\bibfnamefont{D.}~\bibnamefont{Jaksch}},
  \bibinfo{author}{\bibfnamefont{J.~I.} \bibnamefont{Cirac}},
  \bibinfo{author}{\bibfnamefont{P.}~\bibnamefont{Zoller}},
  \bibinfo{author}{\bibfnamefont{S.~L.} \bibnamefont{Rolston}},
  \bibinfo{author}{\bibfnamefont{R.}~\bibnamefont{C\^ot\'e}}, \bibnamefont{and}
  \bibinfo{author}{\bibfnamefont{M.~D.} \bibnamefont{Lukin}},
  \bibinfo{journal}{Phys. Rev. Lett.} \textbf{\bibinfo{volume}{85}},
  \bibinfo{pages}{2208} (\bibinfo{year}{2000}).

\bibitem[{\citenamefont{Laycock et~al.}(2011)\citenamefont{Laycock, Olmos, and
  Lesanovsky}}]{Laycock2011}
\bibinfo{author}{\bibfnamefont{T.}~\bibnamefont{Laycock}},
  \bibinfo{author}{\bibfnamefont{B.}~\bibnamefont{Olmos}}, \bibnamefont{and}
  \bibinfo{author}{\bibfnamefont{I.}~\bibnamefont{Lesanovsky}},
  \bibinfo{journal}{Journal of Physics B: Atomic, Molecular and Optical
  Physics} \textbf{\bibinfo{volume}{44}}, \bibinfo{pages}{184017}
  (\bibinfo{year}{2011}).

\bibitem[{\citenamefont{Lesanovsky}(2012)}]{Lesanovsky2012}
\bibinfo{author}{\bibfnamefont{I.}~\bibnamefont{Lesanovsky}},
  \bibinfo{journal}{Phys. Rev. Lett.} \textbf{\bibinfo{volume}{108}},
  \bibinfo{pages}{105301} (\bibinfo{year}{2012}).

\bibitem[{\citenamefont{Labuhn et~al.}(2016)\citenamefont{Labuhn, Barredo,
  Ravets, Leseleuc, Macri, Lahaye, and Browaeys}}]{Labuhn2016}
\bibinfo{author}{\bibfnamefont{H.}~\bibnamefont{Labuhn}},
  \bibinfo{author}{\bibfnamefont{D.}~\bibnamefont{Barredo}},
  \bibinfo{author}{\bibfnamefont{S.}~\bibnamefont{Ravets}},
  \bibinfo{author}{\bibfnamefont{S.~D.} \bibnamefont{Leseleuc}},
  \bibinfo{author}{\bibfnamefont{T.}~\bibnamefont{Macri}},
  \bibinfo{author}{\bibfnamefont{T.}~\bibnamefont{Lahaye}}, \bibnamefont{and}
  \bibinfo{author}{\bibfnamefont{A.}~\bibnamefont{Browaeys}},
  \bibinfo{journal}{Nature} \textbf{\bibinfo{volume}{534}},
  \bibinfo{pages}{667} (\bibinfo{year}{2016}).

\bibitem[{\citenamefont{de~L{\'e}s{\'e}leuc
  et~al.}(2019)\citenamefont{de~L{\'e}s{\'e}leuc, Lienhard, Scholl, Barredo,
  Weber, Lang, B{\"u}chler, Lahaye, and Browaeys}}]{Leseleuc2019}
\bibinfo{author}{\bibfnamefont{S.}~\bibnamefont{de~L{\'e}s{\'e}leuc}},
  \bibinfo{author}{\bibfnamefont{V.}~\bibnamefont{Lienhard}},
  \bibinfo{author}{\bibfnamefont{P.}~\bibnamefont{Scholl}},
  \bibinfo{author}{\bibfnamefont{D.}~\bibnamefont{Barredo}},
  \bibinfo{author}{\bibfnamefont{S.}~\bibnamefont{Weber}},
  \bibinfo{author}{\bibfnamefont{N.}~\bibnamefont{Lang}},
  \bibinfo{author}{\bibfnamefont{H.~P.} \bibnamefont{B{\"u}chler}},
  \bibinfo{author}{\bibfnamefont{T.}~\bibnamefont{Lahaye}}, \bibnamefont{and}
  \bibinfo{author}{\bibfnamefont{A.}~\bibnamefont{Browaeys}},
  \bibinfo{journal}{Science} \textbf{\bibinfo{volume}{365}},
  \bibinfo{pages}{775} (\bibinfo{year}{2019}), ISSN \bibinfo{issn}{0036-8075}.

\bibitem[{\citenamefont{Samajdar et~al.}(2020)\citenamefont{Samajdar, Ho,
  Pichler, Lukin, and Sachdev}}]{Samajdar2020}
\bibinfo{author}{\bibfnamefont{R.}~\bibnamefont{Samajdar}},
  \bibinfo{author}{\bibfnamefont{W.~W.} \bibnamefont{Ho}},
  \bibinfo{author}{\bibfnamefont{H.}~\bibnamefont{Pichler}},
  \bibinfo{author}{\bibfnamefont{M.~D.} \bibnamefont{Lukin}}, \bibnamefont{and}
  \bibinfo{author}{\bibfnamefont{S.}~\bibnamefont{Sachdev}},
  \bibinfo{journal}{Phys. Rev. Lett.} \textbf{\bibinfo{volume}{124}},
  \bibinfo{pages}{103601} (\bibinfo{year}{2020}).

\bibitem[{\citenamefont{Bernien et~al.}(2017)\citenamefont{Bernien, Schwartz,
  Keesling, Levine, Omran, Pichler, Choi, Zibrov, Endres, Greiner
  et~al.}}]{Bernien2017}
\bibinfo{author}{\bibfnamefont{H.}~\bibnamefont{Bernien}},
  \bibinfo{author}{\bibfnamefont{S.}~\bibnamefont{Schwartz}},
  \bibinfo{author}{\bibfnamefont{A.}~\bibnamefont{Keesling}},
  \bibinfo{author}{\bibfnamefont{H.}~\bibnamefont{Levine}},
  \bibinfo{author}{\bibfnamefont{A.}~\bibnamefont{Omran}},
  \bibinfo{author}{\bibfnamefont{H.}~\bibnamefont{Pichler}},
  \bibinfo{author}{\bibfnamefont{S.}~\bibnamefont{Choi}},
  \bibinfo{author}{\bibfnamefont{A.~S.} \bibnamefont{Zibrov}},
  \bibinfo{author}{\bibfnamefont{M.}~\bibnamefont{Endres}},
  \bibinfo{author}{\bibfnamefont{M.}~\bibnamefont{Greiner}},
  \bibnamefont{et~al.}, \bibinfo{journal}{Nature}
  \textbf{\bibinfo{volume}{551}}, \bibinfo{pages}{579} (\bibinfo{year}{2017}).

\bibitem[{\citenamefont{Keesling et~al.}(2019)\citenamefont{Keesling, Omran,
  Levine, Bernien, Pichler, Choi, Samajdar, Schwartz, Silvi, Sachdev
  et~al.}}]{Keesling2019}
\bibinfo{author}{\bibfnamefont{A.}~\bibnamefont{Keesling}},
  \bibinfo{author}{\bibfnamefont{A.}~\bibnamefont{Omran}},
  \bibinfo{author}{\bibfnamefont{H.}~\bibnamefont{Levine}},
  \bibinfo{author}{\bibfnamefont{H.}~\bibnamefont{Bernien}},
  \bibinfo{author}{\bibfnamefont{H.}~\bibnamefont{Pichler}},
  \bibinfo{author}{\bibfnamefont{S.}~\bibnamefont{Choi}},
  \bibinfo{author}{\bibfnamefont{R.}~\bibnamefont{Samajdar}},
  \bibinfo{author}{\bibfnamefont{S.}~\bibnamefont{Schwartz}},
  \bibinfo{author}{\bibfnamefont{P.}~\bibnamefont{Silvi}},
  \bibinfo{author}{\bibfnamefont{S.}~\bibnamefont{Sachdev}},
  \bibnamefont{et~al.}, \bibinfo{journal}{Nature}
  \textbf{\bibinfo{volume}{568}}, \bibinfo{pages}{207} (\bibinfo{year}{2019}).

\bibitem[{\citenamefont{Turner et~al.}(2019)\citenamefont{Turner, Michailidis,
  Abanin, Serbyn, and Papic}}]{Turner2019}
\bibinfo{author}{\bibfnamefont{C.~J.} \bibnamefont{Turner}},
  \bibinfo{author}{\bibfnamefont{A.~A.} \bibnamefont{Michailidis}},
  \bibinfo{author}{\bibfnamefont{D.~A.} \bibnamefont{Abanin}},
  \bibinfo{author}{\bibfnamefont{M.}~\bibnamefont{Serbyn}}, \bibnamefont{and}
  \bibinfo{author}{\bibfnamefont{Z.}~\bibnamefont{Papic}},
  \bibinfo{journal}{Nature Physics} \textbf{\bibinfo{volume}{14}},
  \bibinfo{pages}{745} (\bibinfo{year}{2019}).

\bibitem[{\citenamefont{Ho et~al.}(2019)\citenamefont{Ho, Choi, Pichler, and
  Lukin}}]{Ho2019}
\bibinfo{author}{\bibfnamefont{W.~W.} \bibnamefont{Ho}},
  \bibinfo{author}{\bibfnamefont{S.}~\bibnamefont{Choi}},
  \bibinfo{author}{\bibfnamefont{H.}~\bibnamefont{Pichler}}, \bibnamefont{and}
  \bibinfo{author}{\bibfnamefont{M.~D.} \bibnamefont{Lukin}},
  \bibinfo{journal}{Phys. Rev. Lett.} \textbf{\bibinfo{volume}{122}},
  \bibinfo{pages}{040603} (\bibinfo{year}{2019}).

\bibitem[{\citenamefont{Prokof'ev et~al.}(1998)\citenamefont{Prokof'ev,
  Svistunov, and Tupitsyn}}]{Prokofev1998}
\bibinfo{author}{\bibfnamefont{N.~V.} \bibnamefont{Prokof'ev}},
  \bibinfo{author}{\bibfnamefont{B.~V.} \bibnamefont{Svistunov}},
  \bibnamefont{and} \bibinfo{author}{\bibfnamefont{I.~S.}
  \bibnamefont{Tupitsyn}}, \bibinfo{journal}{Phys. Lett. A}
  \textbf{\bibinfo{volume}{238}}, \bibinfo{pages}{253} (\bibinfo{year}{1998}).

\bibitem[{\citenamefont{Lesanovsky and Katsura}(2012)}]{Lesanovsky2012b}
\bibinfo{author}{\bibfnamefont{I.}~\bibnamefont{Lesanovsky}} \bibnamefont{and}
  \bibinfo{author}{\bibfnamefont{H.}~\bibnamefont{Katsura}},
  \bibinfo{journal}{Phys. Rev. A} \textbf{\bibinfo{volume}{86}},
  \bibinfo{pages}{041601} (\bibinfo{year}{2012}).

\bibitem[{\citenamefont{Sachdev et~al.}(2002)\citenamefont{Sachdev, Sengupta,
  and Girvin}}]{Sachdev2002}
\bibinfo{author}{\bibfnamefont{S.}~\bibnamefont{Sachdev}},
  \bibinfo{author}{\bibfnamefont{K.}~\bibnamefont{Sengupta}}, \bibnamefont{and}
  \bibinfo{author}{\bibfnamefont{S.~M.} \bibnamefont{Girvin}},
  \bibinfo{journal}{Phys. Rev. B} \textbf{\bibinfo{volume}{66}},
  \bibinfo{pages}{075128} (\bibinfo{year}{2002}).

\bibitem[{\citenamefont{Sengupta et~al.}(2004)\citenamefont{Sengupta, Powell,
  and Sachdev}}]{Sengupta2004}
\bibinfo{author}{\bibfnamefont{K.}~\bibnamefont{Sengupta}},
  \bibinfo{author}{\bibfnamefont{S.}~\bibnamefont{Powell}}, \bibnamefont{and}
  \bibinfo{author}{\bibfnamefont{S.}~\bibnamefont{Sachdev}},
  \bibinfo{journal}{Phys. Rev. A} \textbf{\bibinfo{volume}{69}},
  \bibinfo{pages}{053616} (\bibinfo{year}{2004}).

\bibitem[{\citenamefont{Pielawa et~al.}(2011)\citenamefont{Pielawa, Kitagawa,
  Berg, and Sachdev}}]{Pielawa2011}
\bibinfo{author}{\bibfnamefont{S.}~\bibnamefont{Pielawa}},
  \bibinfo{author}{\bibfnamefont{T.}~\bibnamefont{Kitagawa}},
  \bibinfo{author}{\bibfnamefont{E.}~\bibnamefont{Berg}}, \bibnamefont{and}
  \bibinfo{author}{\bibfnamefont{S.}~\bibnamefont{Sachdev}},
  \bibinfo{journal}{Phys. Rev. B} \textbf{\bibinfo{volume}{83}},
  \bibinfo{pages}{205135} (\bibinfo{year}{2011}).

\bibitem[{\citenamefont{Olmos et~al.}(2012{\natexlab{a}})\citenamefont{Olmos,
  Lesanovsky, and Garrahan}}]{Olmos2012}
\bibinfo{author}{\bibfnamefont{B.}~\bibnamefont{Olmos}},
  \bibinfo{author}{\bibfnamefont{I.}~\bibnamefont{Lesanovsky}},
  \bibnamefont{and} \bibinfo{author}{\bibfnamefont{J.~P.}
  \bibnamefont{Garrahan}}, \bibinfo{journal}{Phys. Rev. Lett.}
  \textbf{\bibinfo{volume}{109}}, \bibinfo{pages}{020403}
  (\bibinfo{year}{2012}{\natexlab{a}}).

\bibitem[{\citenamefont{Olmos et~al.}(2012{\natexlab{b}})\citenamefont{Olmos,
  Gonz{\'{a}}lez-F{\'{e}}rez, Lesanovsky, and Vel{\'{a}}zquez}}]{Olmos2012b}
\bibinfo{author}{\bibfnamefont{B.}~\bibnamefont{Olmos}},
  \bibinfo{author}{\bibfnamefont{R.}~\bibnamefont{Gonz{\'{a}}lez-F{\'{e}}rez}},
  \bibinfo{author}{\bibfnamefont{I.}~\bibnamefont{Lesanovsky}},
  \bibnamefont{and}
  \bibinfo{author}{\bibfnamefont{L.}~\bibnamefont{Vel{\'{a}}zquez}},
  \bibinfo{journal}{Journal of Physics A: Mathematical and Theoretical}
  \textbf{\bibinfo{volume}{45}}, \bibinfo{pages}{325301}
  (\bibinfo{year}{2012}{\natexlab{b}}).

\bibitem[{\citenamefont{Lin and Motrunich}(2019)}]{Lin2019}
\bibinfo{author}{\bibfnamefont{C.-J.} \bibnamefont{Lin}} \bibnamefont{and}
  \bibinfo{author}{\bibfnamefont{O.~I.} \bibnamefont{Motrunich}},
  \bibinfo{journal}{Phys. Rev. Lett.} \textbf{\bibinfo{volume}{122}},
  \bibinfo{pages}{173401} (\bibinfo{year}{2019}).

\bibitem[{\citenamefont{Mark et~al.}(2020)\citenamefont{Mark, Lin, and
  Motrunich}}]{Mark2020}
\bibinfo{author}{\bibfnamefont{D.~K.} \bibnamefont{Mark}},
  \bibinfo{author}{\bibfnamefont{C.-J.} \bibnamefont{Lin}}, \bibnamefont{and}
  \bibinfo{author}{\bibfnamefont{O.~I.} \bibnamefont{Motrunich}},
  \bibinfo{journal}{Phys. Rev. B} \textbf{\bibinfo{volume}{101}},
  \bibinfo{pages}{094308} (\bibinfo{year}{2020}).

\bibitem[{\citenamefont{Michailidis
  et~al.}(2020{\natexlab{a}})\citenamefont{Michailidis, Turner,
  Papi\ifmmode~\acute{c}\else \'{c}\fi{}, Abanin, and
  Serbyn}}]{Michailidis2020}
\bibinfo{author}{\bibfnamefont{A.~A.} \bibnamefont{Michailidis}},
  \bibinfo{author}{\bibfnamefont{C.~J.} \bibnamefont{Turner}},
  \bibinfo{author}{\bibfnamefont{Z.}~\bibnamefont{Papi\ifmmode~\acute{c}\else
  \'{c}\fi{}}}, \bibinfo{author}{\bibfnamefont{D.~A.} \bibnamefont{Abanin}},
  \bibnamefont{and} \bibinfo{author}{\bibfnamefont{M.}~\bibnamefont{Serbyn}},
  \bibinfo{journal}{Phys. Rev. Research} \textbf{\bibinfo{volume}{2}},
  \bibinfo{pages}{022065} (\bibinfo{year}{2020}{\natexlab{a}}).

\bibitem[{\citenamefont{Michailidis
  et~al.}(2020{\natexlab{b}})\citenamefont{Michailidis, Turner,
  Papi\ifmmode~\acute{c}\else \'{c}\fi{}, Abanin, and
  Serbyn}}]{Michailidis2020b}
\bibinfo{author}{\bibfnamefont{A.~A.} \bibnamefont{Michailidis}},
  \bibinfo{author}{\bibfnamefont{C.~J.} \bibnamefont{Turner}},
  \bibinfo{author}{\bibfnamefont{Z.}~\bibnamefont{Papi\ifmmode~\acute{c}\else
  \'{c}\fi{}}}, \bibinfo{author}{\bibfnamefont{D.~A.} \bibnamefont{Abanin}},
  \bibnamefont{and} \bibinfo{author}{\bibfnamefont{M.}~\bibnamefont{Serbyn}},
  \bibinfo{journal}{Phys. Rev. X} \textbf{\bibinfo{volume}{10}},
  \bibinfo{pages}{011055} (\bibinfo{year}{2020}{\natexlab{b}}).

\bibitem[{\citenamefont{Greitemann and Pollet}(2018)}]{Greitemann2018}
\bibinfo{author}{\bibfnamefont{J.}~\bibnamefont{Greitemann}} \bibnamefont{and}
  \bibinfo{author}{\bibfnamefont{L.}~\bibnamefont{Pollet}},
  \bibinfo{journal}{SciPost Phys. Lect. Notes} p.~\bibinfo{pages}{2}
  (\bibinfo{year}{2018}).

\bibitem[{\citenamefont{Domb}(1960)}]{Domb1960}
\bibinfo{author}{\bibfnamefont{C.}~\bibnamefont{Domb}},
  \bibinfo{journal}{Advances in Physics} \textbf{\bibinfo{volume}{9}},
  \bibinfo{pages}{194} (\bibinfo{year}{1960}).

\bibitem[{\citenamefont{Ovchinnikov et~al.}(2003)\citenamefont{Ovchinnikov,
  Dmitriev, Krivnov, and Cheranovskii}}]{Ovchinnikov2003}
\bibinfo{author}{\bibfnamefont{A.~A.} \bibnamefont{Ovchinnikov}},
  \bibinfo{author}{\bibfnamefont{D.~V.} \bibnamefont{Dmitriev}},
  \bibinfo{author}{\bibfnamefont{V.~Y.} \bibnamefont{Krivnov}},
  \bibnamefont{and} \bibinfo{author}{\bibfnamefont{V.~O.}
  \bibnamefont{Cheranovskii}}, \bibinfo{journal}{Phys. Rev. B}
  \textbf{\bibinfo{volume}{68}}, \bibinfo{pages}{214406}
  (\bibinfo{year}{2003}).

\bibitem[{\citenamefont{Sandvik}(2010)}]{Sandvik2010}
\bibinfo{author}{\bibfnamefont{A.~W.} \bibnamefont{Sandvik}},
  \bibinfo{journal}{AIP Conference Proceedings}
  \textbf{\bibinfo{volume}{1297}}, \bibinfo{pages}{135} (\bibinfo{year}{2010}).

\bibitem[{\citenamefont{Samajdar et~al.}(2021)\citenamefont{Samajdar, Ho,
  Pichler, Lukin, and Sachdev}}]{Samajdare2021}
\bibinfo{author}{\bibfnamefont{R.}~\bibnamefont{Samajdar}},
  \bibinfo{author}{\bibfnamefont{W.~W.} \bibnamefont{Ho}},
  \bibinfo{author}{\bibfnamefont{H.}~\bibnamefont{Pichler}},
  \bibinfo{author}{\bibfnamefont{M.~D.} \bibnamefont{Lukin}}, \bibnamefont{and}
  \bibinfo{author}{\bibfnamefont{S.}~\bibnamefont{Sachdev}},
  \bibinfo{journal}{Proceedings of the National Academy of Sciences}
  \textbf{\bibinfo{volume}{118}} (\bibinfo{year}{2021}), ISSN
  \bibinfo{issn}{0027-8424}.

\bibitem[{\citenamefont{{Verresen} et~al.}(2020)\citenamefont{{Verresen},
  {Lukin}, and {Vishwanath}}}]{Verresen2021}
\bibinfo{author}{\bibfnamefont{R.}~\bibnamefont{{Verresen}}},
  \bibinfo{author}{\bibfnamefont{M.~D.} \bibnamefont{{Lukin}}},
  \bibnamefont{and}
  \bibinfo{author}{\bibfnamefont{A.}~\bibnamefont{{Vishwanath}}},
  \bibinfo{journal}{arXiv e-prints} \bibinfo{eid}{arXiv:2011.12310}
  (\bibinfo{year}{2020}), \eprint{2011.12310}.

\end{thebibliography}

\end{document}